\documentclass[10pt,twocolumn]{article}

\begin{document}
\title{\Large{Electromagnetic energy-momentum equation without tensors:\\
a geometric algebra approach}}
\author{Quirino M. Sugon Jr.* and Daniel J. McNamara
\smallskip\\
\small{Ateneo de Manila University, Department of Physics, Loyola Heights, Quezon City, Philippines 1108}\\
\small{*Also at Manila Observatory, Upper Atmosphere Division, Ateneo de Manila University Campus}\\
\small{e-mail: qsugon$@$observatory.ph}}
\date{\small{\today}}
\maketitle

\section*{}
{\small{\textbf{Abstract.} In this paper, we define energy-momentum density as a product of the complex vector electromagnetic field and its complex conjugate.  We derive an equation for the spacetime derivative of the energy-momentum density.  We show that the scalar and vector parts of this equation are the differential conservation laws for energy and momentum, and the imaginary vector part is a relation for the curl of the Poynting vector.  We can show that the spacetime derivative of this energy-momentum equation is a wave equation.  Our formalism is Dirac-Pauli-Hestenes algebra in the framework of Clifford (Geometric) algebra $\mathcal Cl_{4,0}$.}

\section{Introduction}
The conservation laws for electromagnetic energy and momentum are given in Simmons and Guttmann\cite{Simmons and Guttmann 1970 p 174-176} as
\begin{eqnarray}
\label{eq:conservation of energy intro}
-\mathbf E\cdot\mathbf J&=&\frac{\partial U}{\partial t}+\nabla\cdot\mathbf S,\\
\label{eq:conservation of momentum intro}
\rho\mathbf E+\mathbf j\times\mathbf B&=&-\nabla U-\frac{1}{c^2}\frac{\partial\mathbf S}{\partial t}
+\epsilon_0(\nabla\cdot\mathbf E)\mathbf E+(\mathbf E\cdot\nabla)\mathbf E)\nonumber\\
& &+\mu_0(\nabla\cdot\mathbf H)\mathbf H+(\mathbf H\cdot\nabla)\mathbf H).
\end{eqnarray}
where $U$ is the energy density and $\mathbf S$ is the Poynting vector.  Our aim is to unify these two equations.

The standard way to unify Eqs. (\ref{eq:conservation of energy intro}) and (\ref{eq:conservation of momentum intro}) is through tensors as given in Jackson:\cite{Jackson 1975 p 607}
\begin{equation}
\label{eq:D theta is F J}
\partial_\alpha\Theta^{\alpha\beta}=-\frac{1}{c}F^{\beta\lambda}J_\lambda,
\end{equation}
where $\partial_\alpha$ is the spacetime derivative operator, $\Theta^{\alpha\beta}$ is the symmetric stress tensor, $F_{\beta\lambda}$ is the electromagnetic field tensor, and $J_{\lambda}$ is the four-current density.  

Yet, it is not obvious that the tensor equation in Eq. (\ref{eq:D theta is F J}) is equivalent to the vector equations in Eqs. (\ref{eq:conservation of energy intro}) and (\ref{eq:conservation of momentum intro}).  To prove this equivalence is straightforward but tedious.  What we need is a formulation that would enable us to extract the two conservation laws in a single step.

To answer this problem, we propose geometric algebra.\cite{Hestenes 1990,Doran and Lasenby 2003}  This algebra combines the imaginary number $i$ with the dot and cross products of vectors in a single associative vector product:\cite{Hestenes 2003 Feb p 110}
\begin{equation}
\label{eq:Pauli identity intro}
\mathbf a\mathbf b=\mathbf a\cdot\mathbf b+i(\mathbf a\times\mathbf b),
\end{equation}
which looks like the Pauli identity in Quantum Mechanics\cite{Sakurai 1967 p 78} but without the matrices.

One use of the Pauli identity is in the unification of the four Maxwell's equations in Electrodynamics:\cite{Hestenes 2003 Feb p 114, Baylis 1992 p 794}
\begin{equation}
\label{eq:Maxwell equation intro}
(\frac{1}{c}\partial_t+\nabla)(\mathbf E+i\mathbf B)=\rho-\frac{\mathbf j}{c}.
\end{equation}
Separating the scalar, vector, imaginary vector, and imaginary scalar parts of Eq. (\ref{eq:Maxwell equation intro}) yields Gauss's, Ampere's, Faraday's, and magnetic flux continuity laws.  (For comparison, tensor calculus\cite{Landau and Lifshitz 1975 p 67-74} and differential forms\cite{Misner Wheeler and Thorne 1973 p 80-81} can only reduce Maxwell's equations into two.)

So we ask: is there an energy-momentum counterpart to the Maxwell's equation?  

This question was answered by Vold\cite{Vold 1993 p 513} by left-multiplying the Maxwell's equation by the reverse $(\dagger)$ or conjugate of the electromagnetic field $F=\mathbf E+i\mathbf B$, and adding the resulting equation with its reverse:
\begin{equation}
\label{eq:energy momentum equation Vold}
F^\dagger(\dot{\partial}_t+\dot{\nabla})\dot F+\dot {F}^\dagger(\dot{\partial}_t+\dot{\nabla})F=F^\dagger S+S^\dagger F,
\end{equation}
where $S=\rho-\mathbf j/c$ and the overdots determine the direction of differentiation.  The reversion operator, just like the Hermitian conjugation operator in Quantum Mechanics\cite{Bohm 1951 189-190}, changes the order of factors and operators--even the direction of differentiation.

Yet, it is still not obvious that the scalar and vector parts of Eq. (\ref{eq:energy momentum equation Vold}) are indeed Eqs. (\ref{eq:conservation of energy intro}) and (\ref{eq:conservation of momentum intro}).  Showing this result is not trivial, because Eq. (\ref{eq:energy momentum equation Vold}) must first be converted into a form involving the spacetime derivative of the electromagnetic energy-momentum density\cite{Vold 1993 p 512}
\begin{equation}
\label{eq:energy momentum density Hestenes}
\frac{1}{2}FF^\dagger=(\mathbf E+i\mathbf B)(\mathbf E-i\mathbf B)=\frac{1}{2}(|\mathbf E|^2+|\mathbf B|^2)+\mathbf E\times\mathbf B.
\end{equation}
To perform this differentiation properly, we shall use a theorem in Jancewicz\cite{Jancewicz 1988 p 67} for the spatial derivative of a product of two vectors:
\begin{equation}
\label{eq:spatial derivative ab Jancewicz intro}
\nabla(\mathbf a\mathbf b)=(\nabla\mathbf a)\mathbf b-\mathbf a(\nabla\mathbf b)+2(\mathbf a\cdot\nabla)\mathbf b.
\end{equation}
(In Jancewicz, $\mathbf b$ is replaced by $\hat{\mathbf b}=i\mathbf b$; we only factored out the $i$.)  

We shall divide this paper into four sections.  The first is Introduction.  In the second section, we shall review geometric algebra and calculus within the framework of Hestenes's spacetime algebra in spacetime split form via Clifford (Dirac) algebra $\mathcal Cl_{4,0}$.\cite{Hestenes 2003 691-714,Sugon and McNamara 2002 p 297-306}  In the third section, we shall revisit Maxwell's equation and use it to derive the Energy-Momentum equation.  We shall show that the scalar and vector parts of the latter are the two conservations laws, while the imaginary vector part is a relation for the curl of the Poynting vector.  The fourth section is Conclusions.  

\section{Clifford (Geometric) Analysis}

\subsection{Geometric Algebra}

The Clifford (Dirac) algebra $\mathcal Cl_{4,0}$ is generated by four vectors ${\bf e}_0$, ${\bf e}_1$, ${\bf e}_2$, and ${\bf e}_3$ that satisfies the orthonormality relation
\begin{equation}
\label{eq:orthonormality axiom}
{\bf e}_\mu{\bf e}_\nu + {\bf e}_\nu{\bf e}_\mu = 2\delta_{\mu\nu},
\end{equation}
for $\mu,\nu=0,1,2,3$.  That is, ${\bf e}_\mu^2 = 1$ and ${\bf e}_\mu{\bf e}_\nu = -{\bf e}_\nu{\bf e}_\mu$.  We shall refer to ${\bf e}_0$ as the unit temporal vector and to ${\bf e}_1$, ${\bf e}_2$, and ${\bf e}_3$ as the three unit spatial vectors. 

One important subalgebra of $\mathcal Cl_{4,0}$ is the Pauli algebra $\mathcal Cl_{3,0}$.  If ${\bf a}$ and ${\bf b}$ be two vectors spanned by the three unit spatial vectors in $\mathcal Cl_{3,0}$, then by the orthonormality axiom in Eq. (\ref{eq:orthonormality axiom}), we can show that ${\bf a}$ and ${\bf b}$ satisfy the Pauli identity:
\begin{equation}
\label{eq:Pauli identity}
{\bf a}{\bf b} = {\bf a}\cdot{\bf b} + i({\bf a}\times{\bf b}),
\end{equation}
where the imaginary scalar $i={\bf e}_1{\bf e}_2{\bf e}_3$.  In words, the Pauli identity states that the geometric product $\bf ab$ of two vectors is equal to the sum of their dot product ${\bf a}\cdot{\bf b}$ and their imaginary cross product $i(\bf a\times{\bf b})$.

In general, every element $\hat A$ in $\mathcal Cl_{3,0}$ is expressible as a cliffor or a linear combination of a scalar, a vector, an imaginary vector, and an imaginary scalar:
\begin{equation}
\label{ref:spatial cliffor}
\hat A = A_0 + {\bf A}_1 +i{\bf A}_2 + iA_3.
\end{equation}
The spatial inverse or the automorphic grade involution\cite{Baylis 1996 p 4} of $\hat A$ is defined in terms of the unit temporal vector ${\bf e}_0 \equiv\,^\circ$:
\begin{equation}
\label{eq:spatial inversion definition}
\hat A^\circ =\,^\circ\hat A^\dagger. 
\end{equation}
So by the orthonormality axiom in Eq. (\ref{eq:orthonormality axiom}), we have
\begin{equation}
\label{spatial inverse of a cliffor}
\hat A^\dagger = A_0 - {\bf A}_1 +i{\bf A}_2 - iA_3.
\end{equation} 
Notice that the spatial inversion operator changes the sign only of vectors and imaginary scalars.  This differs from the reversion operator which changes sign only of imaginary vectors and imaginary scalars.\cite{Doran and Lasenby 2003 p 39}

\subsection{Geometric Calculus}

\subsubsection{Time Differentiation}

The time derivative of the product of two vector functions ${\bf a}$ and ${\bf b}$ is
\begin{equation}
\label{eq:time derivative of ab}
\frac{\partial}{\partial t}({\bf a}{\bf b}) = \frac{\partial{\bf a}}{\partial t}{\bf b} + {\bf a}\frac{\partial{\bf b}}{\partial t}.
\end{equation}
The scalar and imaginary vector part of this equation are 
\begin{eqnarray}
\label{eq:time derivative of ab scalar}
\frac{\partial}{\partial t}({\bf a}\cdot{\bf b}) &=& \frac{\partial{\bf a}}{\partial t}\cdot{\bf b} + {\bf a}\cdot\frac{\partial{\bf b}}{\partial t},\\
\label{eq:time derivative of ab imaginary vector}
\frac{\partial}{\partial t}({\bf a}\times{\bf b}) &=& \frac{\partial{\bf a}}{\partial t}\times{\bf b} + {\bf a}\times\frac{\partial{\bf b}}{\partial t},
\end{eqnarray}
which are the familiar identities in vector calculus.

\subsubsection{Space Differentiation}

The spatial derivative operator $\nabla$ is defined as
\begin{equation}
\label{eq:del definition}
\nabla=\frac{\partial}{\partial{\bf r}}
={\bf e}_1\frac{\partial}{\partial x_1}+{\bf e}_2\frac{\partial}{\partial x_2}+{\bf e}_3\frac{\partial}{\partial x_3},
\end{equation}
where we used the identity ${\bf e}_k^{-1}={\bf e}_k$.  Because $\nabla$ is a vector operator, we may use the Pauli identity in Eq. (\ref{eq:Pauli identity}) to write
\begin{eqnarray}
\label{eq:spatial derivative of a vector function}
\nabla{\bf a} &=& \nabla\cdot{\bf a}+i(\nabla\times{\bf a}),\\
\label{eq:vector function times spatial derivative}
{\bf a}\nabla &=& {\bf a}\cdot\nabla +i({\bf a}\times\nabla).
\end{eqnarray}
Notice that $\nabla\mathbf a$ is a function, while $\mathbf a\nabla$ is an operator.

For the spatial derivative of the geometric product of two vector functions ${\bf a}$ and ${\bf b}$, we use overdot notation\cite{Doran and Lasenby 2003 p 172-173}:
\begin{equation}
\label{eq:spatial derivative of ab Hestenes overdot}
{\nabla}({\bf a}{\bf b})=({\nabla}{\bf a}){\bf b}+\dot{\nabla}{\bf a}\dot{\bf b},
\end{equation}
where
\begin{equation}
\label{eq:spatial derivative of ab Hestenes overdot second}
\dot{\nabla}{\bf a}\dot{\bf b} = {\bf e}_1{\bf a}\frac{\partial{\bf b}}{\partial x_1}+{\bf e}_2{\bf a}\frac{\partial{\bf b}}{\partial x_2}+{\bf e}_3{\bf a}\frac{\partial{\bf b}}{\partial x_3}. 
\end{equation}
If we employ the orthonormality axiom in Eq. (\ref{eq:orthonormality axiom}), Eq. (\ref{eq:spatial derivative of ab Hestenes overdot second}) becomes
\begin{eqnarray}
\label{eq:spatial derivative of ab Hestenes overdot second alternative}
\dot{\nabla}{\bf a}\dot{\bf b} &=& (a_1{\bf e}_1-a_2{\bf e}_2-a_3{\bf e}_3){\bf e}_1\frac{\partial{\bf b}}{\partial x_1}\nonumber\\
& & +(-a_1{\bf e}_1+a_2{\bf e}_2-a_3{\bf e}_3){\bf e}_2\frac{\partial{\bf b}}{\partial x_3}\nonumber\\
& & +(-a_1{\bf e}_1-a_2{\bf e}_2+a_3{\bf e}_3){\bf e}_3\frac{\partial{\bf b}}{\partial x_3}.
\end{eqnarray}
Adding and subtracting $({\bf a}\cdot{\nabla}){\bf b}$
and rearranging the terms, we get
\begin{equation}
\label{eq:dot del a dot b}
\dot{\nabla}{\bf a}\dot{\bf b}=-{\bf a}{\nabla}{\bf b}+2({\bf a}\cdot{\nabla}){\bf b}.
\end{equation}
Hence,
\begin{equation}
\label{eq:del ab}
\nabla({\bf a}{\bf b}) = (\nabla{\bf a}){\bf b} -{\bf a}(\nabla{\bf b})+2({\bf a}\cdot\nabla){\bf b},
\end{equation}
which is Jancewicz's theorem in Eq. (\ref{eq:spatial derivative ab Jancewicz intro}).  

To verify the correctness of the product rule in Eq. (\ref{eq:del ab}), we separate its vector and imaginary scalar parts:
\begin{eqnarray}
\label{eq:spatial derivative of a product of two vector functions vector part}
\nabla({\bf a}\cdot{\bf b})&-&\nabla\times({\bf a}\times{\bf b})\nonumber\\
&=& (\nabla\cdot{\bf a}){\bf b}-(\nabla\times{\bf a})\times{\bf b}-{\bf a}(\nabla\cdot{\bf b})\nonumber\\
& &+{\bf a}\times(\nabla\times{\bf b})+2({\bf a}\cdot\nabla){\bf b},\\
i\nabla\cdot({\bf a}\times{\bf b}) &=& i(\nabla\times{\bf a})\cdot{\bf b}-i{\bf a}\cdot(\nabla\times{\bf b}).
\end{eqnarray}
The second equation is a familiar identity, while the first can be verified through two other known identities:\cite{Simmons and Guttmann 1970 p 263}
\begin{eqnarray}
\label{eq:del of a dot b}
{\nabla}({\bf a}\cdot{\bf b}) &=& ({\bf a}\cdot{\nabla}){\bf b}+({\bf b}\cdot{\nabla}){\bf a}+{\bf a}\times({\nabla}\times{\bf b})\nonumber\\
& &+{\bf b}\times({\nabla}\times{\bf a}),\\
\label{eq:del cross of a cross b}
{\nabla}\times({\bf a}\times{\bf b}) &=& {\bf a}({\nabla}\cdot{\bf b})-{\bf b}({\nabla}\cdot{\bf a})+({\bf b}\cdot{\nabla}){\bf a}\nonumber\\
& &-({\bf a}\cdot{\nabla}){\bf b}.
\end{eqnarray}

\subsubsection{Spacetime Differentiation}
An event $\hat r^\circ$ is defined by its position $\mathbf r$ and time $t$:
\begin{equation}
\label{eq:event}
\hat r^\circ=(ct+{\bf r})^\circ=\,^\circ(ct-{\bf r})=\,^\circ\hat r^\dagger,
\end{equation}
where $c$ is the speed of light.  The square of the event is the Minkowski metric:
\begin{equation}
\label{eq:r circ squared}
\hat r^\circ\hat r^\circ = \hat r^{\circ\circ}\hat r^\dagger=\hat r\hat r^\dagger=c^2t^2-|{\bf r}|^2.\\
\end{equation}
(In Hestenes's spacetime algebra, the event $x\equiv\hat r^\circ$, so that $x\gamma_0=\hat r^{\circ\circ}=\hat r=ct+\mathbf r$ and $x^2=\hat r^\circ\hat r^\circ$.)

Corresponding to the event $\hat r^\circ$ is the event derivative operator
\begin{equation}
\label{eq:event derivative operator}
\frac{\partial}{\partial\hat r^\circ}=
\,^\circ(\frac{1}{c}\frac{\partial}{\partial t}+{\nabla})=\,^\circ\frac{\partial}{\partial\hat r},
\end{equation}
where we used the identities ${\bf e}_0^{-1}={\bf e}_0$ and $({\bf e}_k{\bf e}_0)^{-1}={\bf e}_0{\bf e}_k$.  We can show that the square of event derivative operator is the d'Alembertian operator for wave equations.\cite{Baylis 1999 p 91}

\section{Classical Electrodynamics}

\subsection{Maxwell's Equation}

In a linear and isotropic medium characterized by the speed of light $c=1/\sqrt{\mu\epsilon}$ and the radiation resistance $\zeta=\sqrt{\mu/\epsilon}$, the Maxwell's equation may be written as
\begin{equation}
\label{eq:Maxwell equation in spacetime}
\frac{\partial\hat E}{\partial\hat r^\circ}=\,^\circ\, \frac{\partial\hat E}{\partial\hat r}=\zeta\hat j^\circ,
\end{equation}
where
\begin{eqnarray}
\label{eq:electromagnetic field}
\hat E &=& {\bf E} + i\zeta{\bf H},\\
\label{eq:spacetime velocity}
\hat j &=& (\rho c + {\bf j}).
\end{eqnarray}
In words, the Maxwell's equation states that the event derivative $\partial/\partial\hat r^\circ$ of the electromagnetic field $\hat E$ is proportional to the event curent density $\hat j^\circ$.\cite{Sugon and McNamara 2002 p 298}

Factoring out the unit temporal vector to the right of Eq. (\ref{eq:Maxwell equation in spacetime}), we get
\begin{equation}
\label{eq:Maxwell equation in space and time}
\frac{\partial\hat E}{\partial\hat r} = \zeta\hat j^\dagger.
\end{equation}
That is,
\begin{equation}
\label{eq:Maxwell equation in space and time expand}
(\frac{1}{c}\frac{\partial}{\partial t} + \nabla)({\bf E} + i\zeta{\bf H}) = \zeta(\rho c - {\bf j}).
\end{equation}
Separating the scalar, vector, imaginary vector, and imaginary scalar parts of Eq. (\ref{eq:Maxwell equation in space and time expand}), we obtain Maxwell's equations:
\begin{eqnarray}
\label{eq:Gauss's law}
\nabla\cdot{\bf E} &=& \zeta\rho c,\\
\label{eq:Ampere's law}
\frac{1}{c}\frac{\partial{\bf E}}{\partial t}-\zeta\nabla\times{\bf H} &=& \zeta{\bf j},\\
\label{eq:Faraday's law}
i(\frac{1}{c}\frac{\partial{\bf H}}{\partial t} + \nabla\times{\bf E}) &=& 0,\\
\label{eq:magnetic flux continuity law}
i\zeta(\nabla\cdot{\bf H}) &=& 0.
\end{eqnarray}

\subsection{Energy-Momentum Equation}

Let us define the event momentum density as
\begin{equation}
\label{eq:event momentum density}
\frac{\hat S^\circ}{c^2}=\frac{1}{c}(U+\frac{{\bf S}}{c})^\circ=-\frac{1}{2c}\epsilon\hat E\hat E^{\dagger\circ}=-\frac{1}{2c}\epsilon\hat E^\circ\hat E.
\end{equation}
where
\begin{eqnarray}
\label{eq:energy density}
U &=& \frac{1}{2}(\epsilon|\mathbf E|^2 + \mu |\mathbf H|^2),\\
{\bf S} &=& {\bf E}\times{\bf H}
\end{eqnarray}
are the energy density and the Poynting vector, respectively.  The square of Eq. (\ref{eq:event momentum density}) is
\begin{equation}
\label{eq:event momentum density squared}
\frac{\hat S^{\circ 2}}{c^4}=\frac{\hat S\hat S^\dagger}{c^4}=\frac{1}{c^2}(U^2-\frac{|{\bf S}|^2}{c^2}).
\end{equation}
If this quantity is zero, then $U =|{\bf S}|/c$, or equivalently, $\mathcal E=|{\bf p}|c$, which is the energy-momentum relation for light-like particles.

The event derivative $\partial/\partial\hat r^\circ$ of the energy-momentum density $\hat S^\circ/{c^2}$ is
\begin{equation}
\label{eq:event derivative of energy-momentum density}
\frac{1}{c^2}\frac{\partial\hat S^\circ}{\partial\hat r^\circ}=-\frac{1}{2c}\epsilon^\circ\frac{\partial}{\partial\hat r}(\hat E\hat E^\dagger)^\circ.
\end{equation}
This leaves us with the problem of differentiating $\hat E\hat E^\dagger$.

By the chain rules in Eqs. (\ref{eq:time derivative of ab}) and (\ref{eq:del ab}), we can show that the temporal and spatial derivatives of $\hat E\hat E^\dagger$ are
\begin{eqnarray}
\label{eq:temporal derivative of E E dagger}
\frac{\partial}{\partial t}(\hat E\hat E^\dagger) &=& \frac{\partial E}{\partial t}\hat E^\dagger+E\,\frac{\partial E^\dagger}{\partial t},\\
\label{eq:spatial derivative of E E dagger}
\nabla(\hat E\hat E^\dagger) &=& (\nabla E)\hat E^\dagger-\hat E\nabla\hat E^\dagger + 2(\hat E\cdot\nabla)\hat E^\dagger,
\end{eqnarray}  
where
\begin{equation}
\label{eq:electromagnetic field dot nabla}
\hat E\cdot\nabla = {\bf E}\cdot\nabla+i\zeta{\bf H}\cdot\nabla.
\end{equation}
Adding Eqs. (\ref{eq:temporal derivative of E E dagger}) and (\ref{eq:spatial derivative of E E dagger}) yields
\begin{equation}
\label{eq:spacetime derivative of E E dagger}
\frac{\partial}{\partial\hat r}(\hat E\hat E^\dagger)= \frac{\partial\hat E}{\partial\hat r}\,\hat E^\dagger+\hat E\,\frac{\partial\hat E^\dagger}{\partial\hat r^\dagger}+2(\hat E\cdot\nabla)\hat E^\dagger.
\end{equation}

If we substitute Eq. (\ref{eq:spacetime derivative of E E dagger}) back to Eq. (\ref{eq:event derivative of energy-momentum density}) and use the Maxwell's equation in Eq. (\ref{eq:Maxwell equation in spacetime}), we get
\begin{equation}
\label{eq:event momentum equation}
\frac{\partial\hat S^\circ}{\partial\hat r^\circ}=-\frac{1}{2}(\hat j^{\circ\circ}\hat E+\,^\circ\hat E\hat j^\circ)-\frac{1}{\zeta}(\hat E\cdot\nabla)\hat E,
\end{equation}
after dividing by $-2\zeta$ and using the spatial inversion property of the temporal vector.  Juxtaposing the two unit temporal vectors to cancel each other, and taking the spatial inverse of the resulting equation, we arrive at
\begin{equation}
\label{eq:energy-momentum equation}
\frac{\partial\hat S}{\partial\hat r} = -\frac{1}{2}(\hat j^\dagger\hat E^\dagger+\hat E\hat j)-\frac{1}{\zeta}(\hat E\cdot\nabla)\hat E^\dagger.
\end{equation}
We shall refer to Eqs. (\ref{eq:event momentum equation}) and (\ref{eq:energy-momentum equation}) as the event-momentum and energy-momentum equations, respectively.  

Expanding Eq. (\ref{eq:energy-momentum equation}),
\begin{eqnarray}
\label{eq:energy-momentum equation spatial inverse expand}
(\frac{1}{c}\frac{\partial}{\partial t}+\nabla)(cU+{\bf S}) &=& -\frac{1}{2}(\rho c-{\bf j})(-{\bf E}+i\zeta{\bf H})\nonumber\\
& &-\frac{1}{2}({\bf E}+i\zeta{\bf H})(\rho c+{\bf j})\nonumber\\
& &-\frac{1}{\zeta}{\bf E}\cdot\nabla(-{\bf E}+i\zeta{\bf H})\nonumber\\
& & -\,i{\bf H}\cdot\nabla(-{\bf E}+i\zeta{\bf H}).
\end{eqnarray}
and separating the scalar, vector, imaginary vector, and imaginary scalar parts, we get
\begin{eqnarray}
\label{eq:conservation of energy}
\frac{\partial U}{\partial t}+\nabla\cdot{\bf S} &=& -{\bf j}\cdot{\bf E},\\
\label{eq:conservation of momentum}
c\nabla U+\frac{1}{c}\frac{\partial{\bf S}}{\partial t} &=& +\frac{1}{\zeta}({\bf E}\cdot\nabla){\bf E}+\zeta({\bf H}\cdot\nabla){\bf H}\nonumber\\
& &-\zeta{\bf j}\times{\bf H},\\
\label{eq:curl of Poynting vector}
i(\nabla\times{\bf S}) &=& i(-\frac{\rho}{\epsilon}{\bf H}-({\bf E}\cdot\nabla){\bf H}+({\bf H}\cdot\nabla){\bf E}),\\
\label{eq:v dot H}
0 &=&-\frac{i}{2}\zeta(-{\bf j}\cdot{\bf H}+{\bf H}\cdot{\bf j}),
\end{eqnarray}
The first equation is the conservation of energy.  The second is the conservation of momentum, which may be converted to that in Eq. (\ref{eq:conservation of momentum intro}) by adding Gauss's law and magnetic flux continuity law.  The third is a relation for the curl of the Poynting vector $\bf S$, which follows from the product rule in Eq. (\ref{eq:del cross of a cross b}).  The fourth is insignificant.

\section{Conclusions}

In this paper, we used Hestenes spacetime algebra in spacetime split form to write down Maxwell's equation, which states that the event derivative oaf the electromagnetic field is proportional to the event current density.  We showed that the four Maxwell's equations may be extracted from this equation by employing the spatial inversion of the temporal vector and Pauli expansion for the geometric product of vectors.

We defined the event momentum density in terms of the product of the electromagnetic field with its spatial inverse.  We used Jancewicz's theorem for the differentiation of a product of two vectors together with the Maxwell's equation to derive the event momentum equation, whose scalar and vector parts are the conservation laws for energy and momentum, and whose imaginary vector part is a relation for the curl of the Poynting vector.  We can show that the event derivative of the event momentum equation would yield its corresponding wave equation.

\section*{Acknowledgments}
This research was supported by the Manila Observatory and by the Vice President for Loyola Schools of the Ateneo de Manila University.

\end{document}